\documentclass{PoS}

\usepackage{latexsym}
\usepackage{graphicx}
\usepackage{epstopdf}
\usepackage{bm}
\title{Nucleon structure from RBC/UKQCD 2+1 flavor DWF dynamical ensembles at a nearly physical pion mass}

\ShortTitle{2+1f DWF Nucleon from RBC/UKQCD}

\author{\speaker{Shigemi Ohta} (RBC and UKQCD Collaborations)\\
     Inst.\ Particle and Nuclear Studies, KEK, Tsukuba, Ibaraki 305-0801, Japan\\
	Department of Particle and Nuclear Physics, Sokendai Graduate University of Advanced Studies, Hayama, Kanagawa 240-0193, Japan\\
	RIKEN BNL Research Center, BNL, Upton, NY 11973, USA\\
	E-mail: \email{shigemi.ohta@kek.jp}}

\abstract{We report the status of nucleon structure calculations on the (2+1)-flavor dynamical domain-wall fermions ensembles with pion masses as low as 180  and 250 MeV on a lattice with about 4.6 fm spatial extent.
A combination of the Iwasaki+dislocation- suppressing-determinant-ratio (I+DSDR)  gauge action and DWF fermion action allows us to generate these ensembles at cutoff of about 1.4 GeV while keeping the residual mass small.
Nucleon source Gaussian smearing has been optimized.
Preliminary nucleon mass estimates are 0.98 and 1.05 GeV.

\vspace{-166mm}\parbox{\textwidth}{\flushright\large\rm \hfill KEK-TH-1419, RBRC-866}\vspace{166mm}
}

\FullConference{The XXVIII International Symposium on Lattice Field Theory, Lattice2010\\
		June 14-19, 2010\\
		Villasimius, Italy}

\begin{document}

\section{Introduction}

RIKEN-BNL-Columbia (RBC) and UKQCD collaborations have been investigating nucleon structure by calculating isovector form factors and some low moments of isovector structure functions \cite{Sasaki:2003jh,Orginos:2005uy,Sasaki:2007gw,Lin:2008uz,Yamazaki:2008py,Yamazaki:2009zq,Aoki:2010xg}.

In our 2+1 flavor dynamical domain-wall fermions (DWF) ensembles with the lattice cutoff of \(a^{-1}=1.73(2)\) GeV and spatial volume of \((2.7{\rm fm})^3\) \cite{Allton:2008pn}, we found surprisingly strong volume dependence in isovector axialvector-current form factors \cite{Yamazaki:2008py,Yamazaki:2009zq}: lattice spatial extent of five pion Compton wave lengths or larger seems necessary to reproduce these quantities.
This provided the first concrete evidence of virtual pion cloud surrounding nucleon.
While the vector-current form factors do not show such strong volume dependence, they fail to reproduce the experimental values as the pion mass in these calculations stayed much higher, at about 330 MeV, than the experimental value of about 140.
It would be very interesting to do similar lattice-QCD calculations at lower pion mass and large spatial volume.

On the other hand for low moments of isovector structure functions such as quark momentum or helicity fractions, our calculation at the lightest pion mass of about 330 MeV showed encouraging trending toward the experimental values \cite{Aoki:2010xg}:
calculations at lighter quark mass values are necessary to confirm if this interesting trend is a real physics effect.

Fortunately, the new 2+1 flavor dynamical DWF ensembles we started to generate last year at the lower cutoff of \(a^{-1}=1.368(7)\) GeV \cite{BobLat2010} are ideal for such a study of nucleon structure at light quark mass and large spatial volume:
the pion mass is about 180 and 250 MeV for the two ensembles and the spatial extent with 32 lattice units is physically about 4.6 fm. 
Here we report the current status of the nucleon structure calculations with these ensembles.

\section{Form factors}

The isovector form factors are conveniently summarized in neutron \(\beta\)-decay transition matrix elements,
\begin{equation}
\langle p| V^+_\mu(x) | n \rangle = \bar{u}_p \left[\gamma_\mu
F_V(q^2) + \frac{\sigma_{\mu \lambda}q_{\lambda}}{2m_N} {F_T(q^2)} \right]
u_n e^{iq\cdot x},
\end{equation}
\begin{equation}
\langle p| A^+_\mu(x) | n \rangle = \bar{u}_p
            \left[\gamma_\mu  \gamma_5 F_{A}(q^2)
             +i q_\mu \gamma_5 {F_{P}(q^2)} \right]  u_n e^{iq\cdot x}.
\end{equation}

The vector-current form factors, the vector,  \(F_V=F_1\), and tensor, \(F_T=F_2\), are identical to the isovector components of the nucleon electromagnetic form factors, and are sometimes represented as electric, \(\displaystyle G_E(q^2)=F_1-\frac{q^2}{4m_N^2}F_2\), and magnetic, \(G_M=F_1+F_2\), form factors.
From these we deduce such quantities as mean-squared charge radius and magnetic moment.
These form factors have important implications in atomic physics as they determine nuclei interaction with photon and electron.
They are under very active experimental studies: indeed a recent experiment \cite{Pohl:2010zz} is in systematic disagreement with earlier ones about proton mean-squared charge radius.
Our lattice-QCD numerical calculations of these quantities may help resolve this discrepancy.

The vector form factor at zero momentum transfer is also related to the Fermi constant and Cabibbo mixing: \(\displaystyle g_V=F_V(0) = G_{\rm Fermi} \cos\theta_{\rm Cabibbo}\).
The axial form factor at zero momentum transfer, the axial charge deviates from unity in units of \(g_V\) because of  QCD correction, \(\displaystyle g_A=F_A(0)=1.2694(28)g_V\) \cite{PDG2010}.
Whether the lattice-QCD numerical calculation can reproduce this quantity is an interesting test of the method.
The vector, \(g_V\),  and axial, \(g_A\),  charges determine the \(\beta\)-decay life time of neutron.
The latter also dominates pion-nucleon interaction through the Goldberger-Treiman relation, \(\displaystyle m_Ng_A \propto f_\pi g_{\pi NN}\).
Thus these quantities determine primordial and stellar nucleosyntheses and are important in explaining abundance of elements.

To calculate these form factors in lattice QCD, we usen ratio of two- and three-point correlators such as \(\displaystyle \frac{C_{\rm 3pt}^{\Gamma,O}(t_{\rm sink}, t)}{C_{\rm 2pt}(t_{\rm sink})}\) with
\begin{equation}
C_{\rm 2pt}(t_{\rm sink}) =
\sum_{\alpha, \beta}
\left(
\frac{1+\gamma_t}{2}
\right)_{\alpha\beta}
\langle
N_\beta(t_{\rm sink})\bar{N}_\alpha(0)
\rangle,
\end{equation}
\begin{equation}
C_{\rm 3pt}^{\Gamma, O}(t_{\rm sink}, t) =
\sum_{\alpha, \beta}
\Gamma_{\alpha\beta}
\langle
N_\beta(t_{\rm sink})O(t)\bar{N}_\alpha(0)
\rangle,
\end{equation}
where \(N\) denotes an appropriate nucleon operator, \(O\) the relevant current operator, and \(\Gamma\) an appropriate spin or momentum-transfer projections.
In this study we choose a standard nucleon operator, \(N=\epsilon_{abc}(u_a^T C \gamma_5 d_b) u_c\).
For details of the spin or momentum-transfer projectors, \(\Gamma\), we refer our earlier publications such as ref.\ \cite{Lin:2008uz}.
The ratio gives a plateau in \(0< t< t_{\rm sink}\) for a lattice bare value \(\langle O\rangle\) for the relevant observable.
We will discuss later how to optimize the sink position in time, \(t_{\rm sink}\), to minimize systematics from excited-state contamination.

Most recently we calculated all the four form factors using our 2+1 flavor dynamical DWF ensembles at lattice cutoff of \(a^{-1}=1.73(2)\) GeV and spatial extent of 2.7 fm \cite{Yamazaki:2008py,Yamazaki:2009zq}: at the lightest pion mass of about 330 MeV we found huge finite-volume corrections in the axialvector-current form factors.
While the vector-current form factors do not suffer from such a finite-volume correction, their values generally deviate from the experiments very likely because our degenerate up and down quark mass is still far from reality.
Thus we like to calculate these quantities at more realistically light mass and at the same time in sufficiently large spatial volume.

\section{Moments of structure functions}

The structure functions are measured in lepton deep-inelastic scattering off nucleon.
The cross section is factorized in terms of leptonic and hadronic tensors, \(\displaystyle \propto \frac{\alpha^2}{(q^2)^2}  l^{\mu\nu} W_{\mu\nu}\), and because we know the leptonic sufficiently well, \(\displaystyle l_{\mu\nu} = 2 (k_\mu k'_\nu + k'_\mu k_\nu - \frac{1}{2}Q^2g_{\mu\nu})\), we deduce the hadronic tensor, \(W^{\mu\nu}\),  from which we extract the structure functions: spin-unpolarized structure functions, \(F_1\) and \(F_2\), are obtained from symmetrized hadronic tensor,
\begin{equation}
W^{\{\mu\nu\}}(x,Q^2) =
\left( -g^{\mu\nu} + \frac{q^\mu q^\nu}{q^2}\right)
 {F_1(x,Q^2)} \nonumber + 
\left(P^\mu-\frac{\nu}{q^2}q^\mu\right)\left(P^\nu-\frac{\nu}{q^2}q^\nu\right)
\frac{F_2(x,Q^2)}{\nu}, 
\end{equation}
and spin-polarized ones, \(g_1\) and \(g_2\), are from antisymmetrized tensor,
\begin{equation}
W^{[\mu\nu]}(x,Q^2) =
i\epsilon^{\mu\nu\rho\sigma} q_\rho
\left(\frac{S_\sigma}{\nu}({g_1(x,Q^2)}+
                           {g_2(x,Q^2)}) - 
\frac{q\cdot S P_\sigma}{\nu^2}{g_2(x,Q^2)} \right),
\end{equation}
with \(\nu = q\cdot P\), \(S^2 = -M^2\), \(x=Q^2/2\nu\), and \(Q^2=|q^2|\).

What are accessible for lattice-QCD calculations are moments of the structure functions, such as
\begin{equation}
2 \int_0^1 dx x^{n-1} F_1(x, Q^2)
= \sum_{q=u, d} c^{(q)}_{1,n}(\mu^2/Q^2, g(\mu)) \langle x^n\rangle_q(\mu) +O(1/Q^2),
\end{equation}
where \(c_{1,n}\) are perturbative Wilson coefficients in the operator product expansion, and \({\langle x^n \rangle_{q}(\mu)}\) are the moments we seek.
The other unpolarized structure function, \(F_2\), is similarly expressed in terms of the same  \({\langle x^n \rangle_{q}(\mu)}\) moments but with different Wilson coeeficients \(c_{2,n}\).
From polarized structure functions \(g_1\) and \(g_2\) we obtain \({\langle x^n \rangle_{\Delta q}(\mu)}\) and \(d_n(\mu)\) moments.
In addition to these moments of DIS structure functions, we calculate tensor charge, \(\langle 1 \rangle_{\delta q} = \langle P, S | \bar{\psi} i \gamma_5 \sigma_{\mu\nu} \psi | P, S\rangle\) which may be measured by polarized Drell-Yan and RHIC Spin experiments.

As the operator product expansion indicates, these moments need be renormalized before we can compare them with experiments.
Good continuum-like flavor and chiral symmetries of domain-wall fermions makes this non-perturbative renormalization procedure feasible.
We follow Rome-Southampton procedure \cite{Dawson:1997ic,Blum:2001sr}.

These moments can be calculated in lattice QCD as forward matrix elements of certain local operators, such as \(\displaystyle \overline{q} \left[\gamma_4 \stackrel{\leftrightarrow}{D_4} - \frac{1}{3}\sum_k \gamma_k \stackrel{\leftrightarrow}{D_k} \right] q\).
Such matrix elements are obtained from lattice QCD as ratios of two- and three-point correlators just like in form factor calculations.
We calculate those moments that do not require finite momentum transfer, namely \(\langle x\rangle_q\), \(\langle 1\rangle_{\Delta q}\), \(\langle x \rangle_{\Delta q}\), and \(d_1\).
For details we refer our earlier publication \cite{Lin:2008uz,Aoki:2010xg} and references cited therein.

We calculated these low moments using the same 1.7-GeV, 2.7-fm, (2+1)-flavor dynamical DWF ensembles \cite{Aoki:2010xg}.
In contrast to the axialvector-current form factors, we have not seen any sign of finite-volume correction in any of these low moments.
Furthermore the results for momentum, \(\langle x\rangle_{u-d}\), and helicity, \(\langle x \rangle_{\Delta u - \Delta d}\), fractions obtained from a smaller, 1.8-fm, box at the same cutoff, agree with the larger box results.
And at the lightest pion mass of 330 MeV these two quantities show interesting trending toward the experiments.
Thus it would be interesting to study these moments at lighter mass.

\section{Iwasaki+DSDR ensembles}

We have been generating a new series of (2+1)-flavor dynamical DWF ensembles in a larger physical volume of about \((4.6 {\rm fm})^3\) \cite{BobLat2010}.
This was made possible by moving to a coarser lattice cutoff of \(a^{-1}=1.368(7)\) GeV.
If we were to use such a coarse cutoff with Iwasaki gauge action we used for earlier studies, then we would have suffered from too large residual violation of the chiral symmetry.
Thus we changed our gauge action by multiplying it with a ratio of Wilson-Dirac fermion determinants \cite{Vranas:1999rz,Vranas:2006zk,Renfrew:2009wu} that adequately suppress gauge dislocation that would result in unacceptably large residual violation of chiral symmetry while maintaining total gauge homotopy distribution comfortable for ergodicity of Monte Carlo gauge update algorithm.

Two ensembles are being generated, one with degenerate up and down quark mass of 0.001 lattice units and another at 0.0042.
The strange quark mass is set at almost exactly physical point, at 0.0045 lattice units.
With the observed residual mass of about 0.002 lattice units, the former ensemble corresponds to about 180 MeV pion mass, and the latter to about 250 MeV.
We have accumulated about 1,100 hybrid Monte Carlo time units for the former and almost 1,900 for the latter.
Considering the thermalization time of 500 for the former and 600 for the latter, and the observable calculation interval of 8 time units, there are about 75 usable configuration for the former and 160 for the latter.
At the time of the conference we had analyzed about 30 from the former and 20 from the latter.
The statistics for the former stays the same while we have increased to about 40 for the latter now.

\section{Lattice systematic errors}

\begin{figure}[t]
\begin{center}
\includegraphics[width=.495\textwidth]{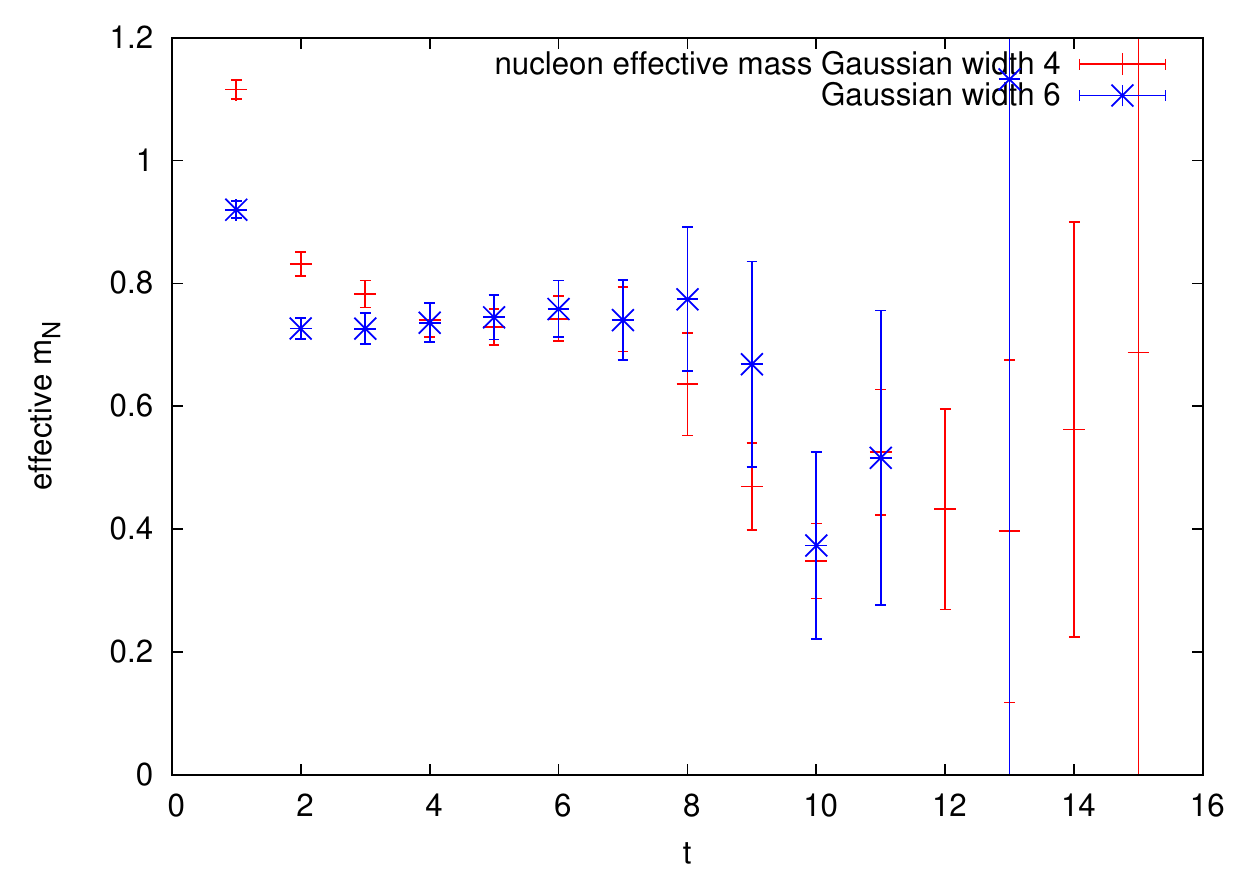}
\includegraphics[width=.495\textwidth]{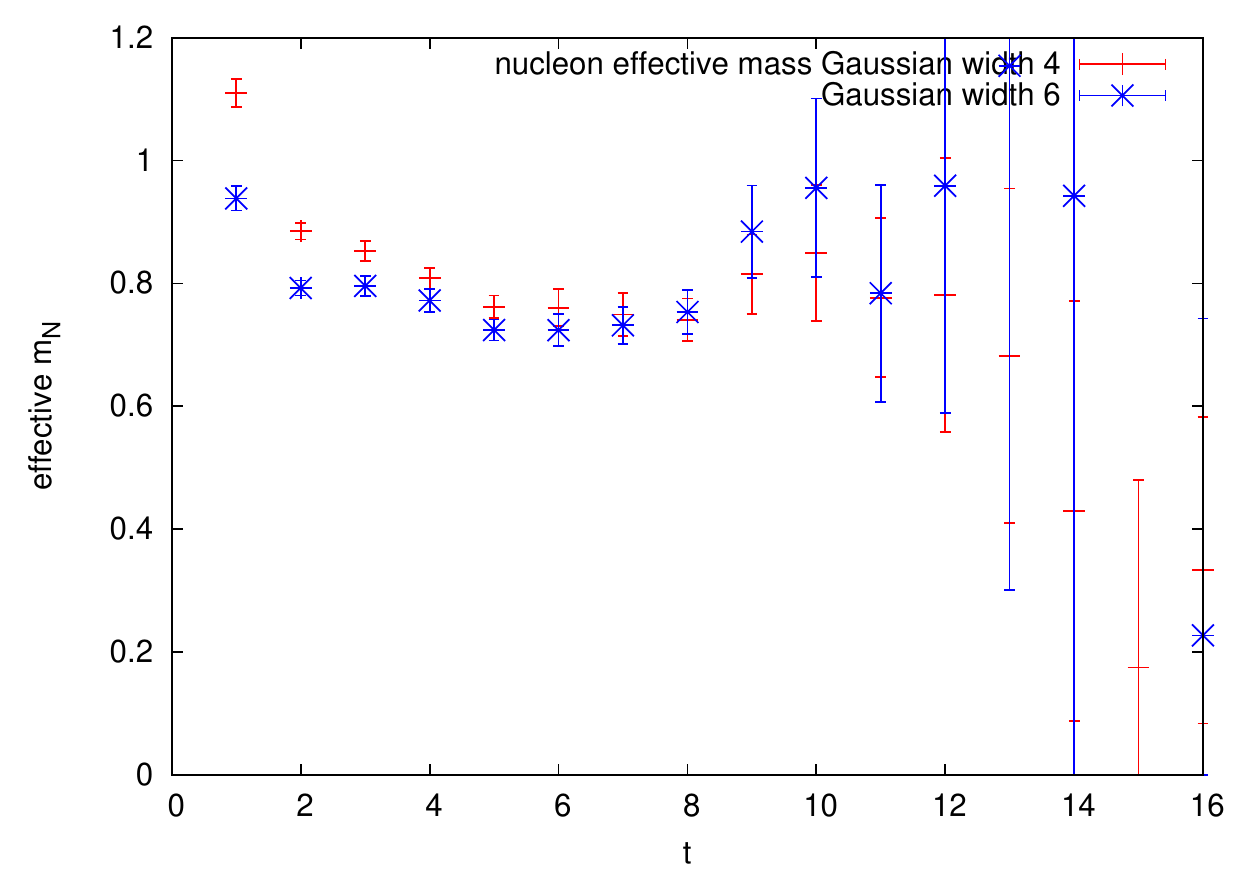}
\end{center}
\caption{Preliminary nucleon effective mass with Gaussian smeared source of width 4 and 6 lattice units and point sink: the light, \(m_\pi=180\) MeV, (left) and heavy, 250 MeV, (right).
The lighter-ensemble results are from about 30 configurations each, while the heavier is from about 40.}
\label{fig:nucleon_effective_mass}
\end{figure}
From our earlier studies we have identified two important sources of systematic errors \cite{Ohta:2009uy}: 1) time separation between nucleon source at time \(t=0\) and sink at \(t_{\rm sink}\), and 2) spatial volume.
In the above we have already explained how the latter is a serious problem for axialvector-current form factors but not for the other nucleon observables we are interested in.
For the axialvector-current form factors the lattice spatial extent of five or more pion Compton wave lengths would be sufficient.
Since our lightest pion mass now is about 180 MeV and the Compton wave length is about 1.1 fm, the spatial extent of about 4.6 fm may not be sufficiently large.
But it is sufficient for the heavier pion mass of about 250 MeV.
Thus we should at least be able to study the finite-volume correction in more detail.

The source-sink time separation concerns excited-state contamination.
No matter how we choose the nucleon operator, it would create some excited-state components at the source, \(t=0\).
Though such contamination should decay  more quickly in time than the ground state, it can still contaminate our calculation if we set the sink too close to the source in time.
In our RBC 2-flavor dynamical DWF study \cite{Lin:2008uz} we have demonstrated that such excited-state contamination can be serious in some of the nucleon observables we are interested in: results for the isovector momentum fraction, \(\langle x \rangle_{u-d}\) shifted beyond statistical error when we changed the source-sink separation from about 1.1 fm to 1.3 fm.
Of course the optimum separation depends on the details of the nucleon operator and the nucleon mass spectrum on the given gauge ensemble.
So prior to the three-point correlator calculations it is imperative to optimize the combination of nucleon operator and source-sink separation.

We chose to use a standard nucleon operator, \(N=\epsilon_{abc}(u_a^T C \gamma_5 d_b) u_c\), and optimize its Gaussian smearing \cite{Alexandrou:1992ti,Berruto:2005hg}.
For both the light, \(m_\pi=180\) MeV and 250 MeV ensembles we have compared the Gaussian width of 4 and 6 lattice units.
In Fig.\ \ref{fig:nucleon_effective_mass} we present the nucleon effective mass with point sink obtained from the current preliminary statistics about 30 for the light and 40 for the heavy ensembles.
In both ensembles the wider, width-6, results settle on a plateau more quickly, and the narrower width-4 results merge with them.
From these observations we conclude the wider width-6 Gaussian smearing is sufficient for our study.

With the present preliminary statistics of 30 or 40, the plateaux decay beyond 8 or 10 lattice spacings or about 1.1 or 1.4 fm.
But since we have accumulated 75 configurations for the light and 160 for the heavy ensembles, and since we are adding two more source positions in time at \(t=16\) and 48 in addition to the current 0 and 32, we will soon quadruple our statistics: source-sink separation of about 1.4 fm should be available.

\section{Summary}

\begin{figure}[t]
\begin{center}
\includegraphics[width=0.7\textwidth]{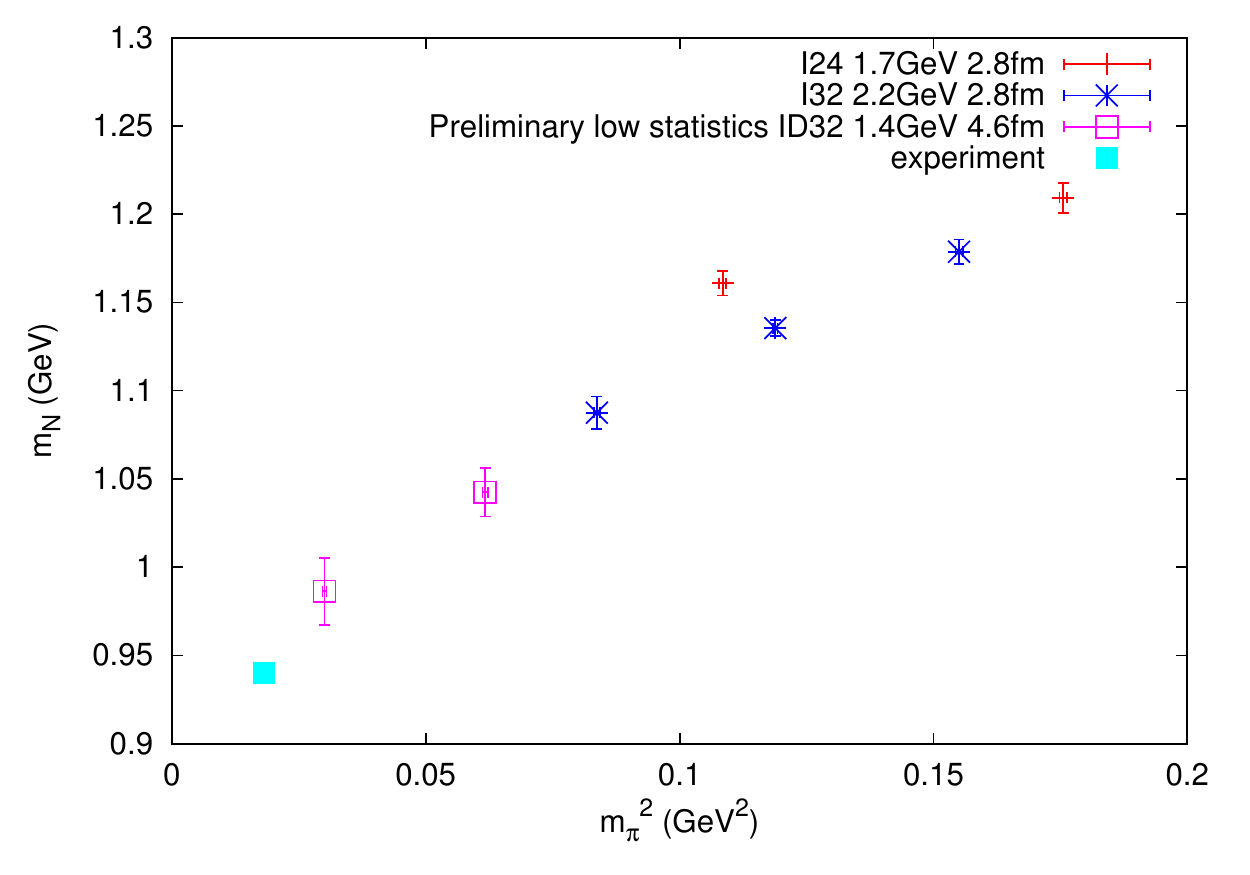}
\end{center}
\caption{Nucleon mass, \(m_N\), plotted against pion mass squared, \(m_\pi^2\), from RBC+UKQCD (2+1)-flavor dynamical DWF ensembles.
The latest and lightest two points at \(m_\pi=250\) and 180 MeV, though preliminary, seem trending downward toward the experiment.}
\label{fig:mpi2mN}
\end{figure}

We are starting our nucleon structure calculations using the new series of RBC/UKQCD joint (2+1)-flavor dynamical DWF ensembles \cite{BobLat2010} with pion mass as low as 180 and 250 MeV and lattice spatial extent as large as 4.6 fm.
As is shown in Fig.\  \ref{fig:mpi2mN}, preliminary estimate for nucleon mass is 0.721(13) lattice units or 0.98 GeV when pion mass is 180 MeV, and 0.763(10) or 1.05 GeV when pion mass is 250 MeV.
They seem trending down toward the experiment more strongly than our previous studies at heavier pion mass.
We have optimized our Gaussian smearing of nucleon source, and will be reporting results for isovector form factors and some low moments of isovector structure functions in the near future.

\section*{Acknowlegment}

SO thanks his collaborators in RBC and UKQCD Collaborations, especially Meifeng Lin, Yasumichi Aoki, Tom Blum, Chris Dawson, Taku Izubuchi, Chulwoo Jung, Shoichi Sasaki and Takeshi Yamazaki.
RIKEN, Brookhaven National Laboratory, the U.S.\ Department of Energy, University of Edinburgh, and the U.K.\  PPARC provided  facilities essential for the completion of this work.
The Iwasaki+DSDR ensembles are being generated at Argonne National Laboratory  Leadership Class Facility (ALCF.)
The nucleon two- and three-point correlators are being calculated at RIKEN Integrated Cluster of Clusters (RICC.)

\end{document}